\documentclass[showpacs,preprintnumbers,12pt]{revtex4}
\usepackage{graphicx}
\usepackage{dcolumn}
\makeatletter
\newcommand{\doublespacing}{\let\CS=\@currsize\renewcommand{\baselinestretch}
{2.0}\tiny\CS}
\input{epsf}

\begin{document}
\title{ On  atomic analogue of Landau quantization }

\author{B. Basu}
\email{banasri@isical.ac.in, Fax:91+(033)2577-3026}
\affiliation{Physics and Applied Mathematics Unit\\
 Indian Statistical Institute\\
 Kolkata-700108 }
\author{S. Dhar}
\email{sarmi_30@rediffmail.com}
\affiliation{
 Bankura Sammilani College, Bankura,W.B., India}
 \author{S. Chatterjee} \email{sankar_2050@yahoo.co.in}
 \affiliation{Bispara High School, Chinsurah, Hooghly W.B., India}

\vspace*{4cm}

\begin{abstract}
\begin{center}
{\bf Abstract}
\end{center}
We have studied the physics of atoms with permanent electric dipole
moment and non vanishing magnetic moment interacting with an
electric field and inhomogeneous magnetic field. This system can be
demonstrated as the atomic analogue of Landau quantization of
charged particles in a uniform magnetic field. This  Landau-like atomic
problem is also studied with space-space noncommutative coordinates.
\end{abstract}

\maketitle

\newpage
\section{Introduction}
A charged particle in two dimensions under a transverse uniform magnetic field gives rise to the well known Landau quantization \cite{1} where a discrete energy spectrum is observed.
 This can be shown as the simplest model to
explain the integer quantum Hall effect (QHE)\cite{mac}. Although the Hall effect of neutral atoms was predicted long ago \cite{kas}, the possibility of an analogue of the Hall effect in Bose Einstein condensates was studied recently by Paredes et. al.
 \cite{par,par1}. Inspired by this work \cite{par,par1},
  the Aharonov-Casher interaction
 was used to generate an analogue of Landau levels in a system of neutral atoms \cite{eric}.
 This is a step towards an atomic QHE using electric fields realizable in Bose Einstein condensates.
  The quantum dynamics of cold atoms in presence of external fields
 is also of current interest \cite{opt1,opt3,dien,duan,mura,spin,sp1}. Recently, the idea of Ericsson and Sj\"{o}qvist
\cite{eric} has been extended to
 investigate the  quantum
dynamics of a neutral particle in the presence of an external
electromagnetic field \cite{nas1,nas2}. It is  shown that in a specific field-dipole
configuration, a quantization similar to Landau quantization may be
obtained \cite{nas1,nas2}.
  These results motivated us to study the
quantum dynamics of an atom in presence of external electric
 and inhomogeneous magnetic fields.

 We have organised the paper in the following way. In sec.II the Landau
like quantiation of a neutral atom in a specific configuration of external fields is studied. In sec.III, we have studied the system in symmetric gauge and derived the eigenfunction and the energy spectrum. Sec.IV is devoted to investigate the system in space-space noncommutative coordinates. Finally, we have discussed an overview of our work in section V.

\section{Landau-like quantization of neutral particles}

In the nonrelativistic limit, the Hamiltonian for a neutral particle
(of mass $m$) that possesses permanent electric dipole moment ${\bf
d_e}$ and nonvanishing magnetic moment ${\bf d_m}$ in  presence
of external electric and magnetic fields is given by \cite{pass}
\begin{equation}\label{a1}
H=-\frac{1}{2m}\left( {\nabla}-{\bf (d_{m}\times E)+(d_{e}\times
B)}\right)^2 -\frac{d_{m}}{2m} {\bf\nabla}.{\bf E} +
\frac{d_{e}}{2m}{\bf\nabla}.{\bf B}
\end{equation}
In this case, the terms of order $(E^2~ \mathrm{and} ~B^2)$ are
neglected, and the electric and magnetic dipoles are aligned in the
z- direction. It is also assumed that the fields generated by the
source are radially distributed in the space. The  momentum operator
may be written as,
\begin{equation}\label{a2}
{ k_{i}}=mv_{i}={ p_{i}}- {\bf (d_{m} \times E)}_{i} + {\bf (d_{e}
\times B)}_{i}
\end{equation}

We may now consider a different set up. The neutral atoms, with a
permanent electric dipole moment $d_e$ and nonvanishing magnetic
moment $d_m$ are made to interact with an electric field ${\bf E}$
and a specific type of magnetic field ${\bf B}$. Let an
inhomogeneous magnetic field be generated in the neighbourhood of
the dipole. In particular,  a non zero gradient of the magnetic
field component perpendicular to the surface S (xy plane), varying
in the direction of the dipole  is subjected. In the nonrelativistic
limit, the interaction between the atom and this type of external
field may be described by the Hamiltonian as ($c=\hbar=1$)
\begin{equation}\label{h1}
H= \frac{{\bf \Pi}^2}{2m} +\frac{d_m }{2m}{\bf \nabla}.{\bf E}+
\frac{d_e }{2m} |(\bf n^\prime . \nabla){\bf B}|
\end{equation}
where ${\bf n}$ is the unit vector in the direction of ${\bf d_{m}}$
and ${\bf n^{'}}$ is the direction of electric dipole moment ${\bf
d_{e}}$. The minimal coupling of the particle with the external
field is obtained by substituting its momentum as
\begin{equation}\label{a3}
\begin{array}{ccl}
{\bf\Pi}&=&-i{\bf\nabla}-d_m ({\bf n}\times{\bf E})-d_{e}({\bf
n^{'}}.{\bf\nabla}){\bf A}\\
&=&{\bf p}-{\bf A}_{eff}
\end{array}
\end{equation}
where
${{\bf A}_{eff}}=d_m({\bf n}\times{\bf E})+d_e({\bf
n^{'}}.{\bf\nabla}){\bf A}$ is defined as the effective vector
potential. A careful analysis of eqn.(\ref{h1}) shows that the
particles can give rise to different phenomena that generate quantum
phases while undergoing a cyclic trajectory. The second term is the
origin of Aharanov Casher effect, while the last term is due to $differential$
  $Aharanov$ $Bohm$ $effect$ (DABE)\cite{5,pach}. Actually, this term involves the coupling of the electric dipole moment to a differential form of the vector potential and so may be defined as DABE \cite{5,pach}.

 It is noted that the momentum ${\bf \Pi}$ follows the standard algebra:
\begin{equation}
[\Pi_i,\Pi_j]=i\epsilon_{ijk}(B_{eff})_k
\end{equation}
where the effective field strength ${\bf B}_{eff}$ is written as
\begin{equation}\label{a5}
\begin{array}{lcl}
{\bf B}_{eff}&=&{\bf \nabla}\times{{\bf A}_{eff}}\\
&=& {\bf \nabla}\times[d_m({\bf n}\times{\bf E})+d_e({\bf
n^{'}}.{\bf \nabla}){\bf A}]\\
\end{array}
\end{equation}

We can impose the  specific field dipole configurations:\\
(i) electric dipole moment in the x- direction i.e.${\bf n}=(0,0,1)$\\
(ii)magnetic dipole moment along the z- direction i.e.  ${\bf n^{'}}=(1,0,0)$\\
(iii)Electric field on the xy surface: ${\bf E}=(E_{x}(x,y),E_{y}(x,y),0)$  with  ${\partial_t E}=0$ and ${\bf\nabla\times\bf E=0}$
and \\
(iv) with ${\bf B}=(0,0,B_{z}(x,y))$ so that
${\bf\nabla.\bf B=0}$ and $ {\bf\nabla\times\bf B=0}$.

 A simple algebra and with the help of Gauss' law, eqn.(\ref{a5}) becomes
\begin{equation}\label{e}
\begin{array}{lcl}
{{\bf B}_{eff}}&=&\displaystyle{d_m[{\bf \nabla}\times({\bf
n}\times{\bf E})]+d_e[{\bf \nabla}\times({\bf n^{'}}.{\bf
\nabla}){\bf A}]}
\\
&&\\ &=&d_m[ { \bf n}({\bf \nabla} . {\bf E})+ {\bf \nabla}({\bf
n}.{\bf E})]+d_e[{\bf \nabla}\times({\bf
B}\times{\bf n}^{'})+{\bf \nabla}\times{\bf \nabla}({\bf n}^{'}.{\bf A})]\\
&&\\&=&d_m[ { \bf n}({\bf \nabla} . {\bf E})]+d_e[{\bf
\nabla}\times({\bf B}\times{\bf n}^{'})]\\&&\\ &=&d_m{\bf
n}\frac{\rho_{0}}{\epsilon_{0}}+d_e(\bf n^{'}.\bf\nabla)\bf B
\end{array}
\end{equation}

It may be noted that a non zero effective phase
 \begin{equation}\label{a}
\phi=\int_{s}[d_m{\bf n}\frac{\rho_{0}}{\epsilon_{0}}+d_e({\bf
n}.{\bf\nabla}){\bf B}].d{\bf S}
\end{equation}
is produced if the z-component of magnetic field has a non-vanishing
gradient along the direction of $d_{e}$ \cite{pach}, $i.e$ a non-zero
effective phase exists when $\frac{\partial B_{z}}{\partial x}\neq 0$

The kinetic part of the Hamiltonian operator satisfies the relation
\begin{equation}\label{e4}
[\Pi_{x},\Pi_{y}]=im\omega_{eff}=\frac{i\sigma}{l_{eff}^2}
\end{equation}
where the  cyclotron frequency is given by
\begin{equation}\label{e5}
\omega_{eff}=\frac{\sigma{|d_{m}\frac{\rho_{0}}{\epsilon_0}+d_{e}(\frac{\partial
B_{z}}{\partial x})|}}{m}
\end{equation}
and the sign $\sigma=\pm 1$ denotes the direction of revolution. In this specific
problem, the magnetic length is defined as
\begin{equation}\label{e6}
l_{eff}=[\frac{1}{\mid{d_{m}\frac{\rho_{0}}{\epsilon_0}}+d_{e}{(\frac{\partial
B_{z}}{\partial x})\mid}}]^{\frac{1}{2}}
\end{equation}

Defining the annihilation and creation operator as
\begin{equation}\label{e7}
a_{eff}={\frac{1}{\sqrt{
2m\mid\omega_{eff}\mid}}}(\Pi_{x}+i\sigma\Pi_{y})~,~
a_{eff}^{\dag}={\frac{1}{\sqrt{
2m\mid\omega_{eff}\mid}}}(\Pi_{x}-i\sigma\Pi_{y})
\end{equation}
with $[a_{eff},a_{eff}^{\dag}]=1$, we can write the Hamiltonian as
\begin{equation}\label{e8}
H=(a_{eff}^{\dag}a_{eff}+\frac{1}{2}(1+\sigma))\mid\omega_{eff}\mid+\frac{p_{z}^2}{2m}
\end{equation}
where we have used the relation $\frac{d_m }{2m}{\bf \nabla}.{\bf
E}+ \frac{d_e }{2m} ({\bf n}^\prime . \nabla){\bf
B}=\frac{1}{2}\omega_{eff}$

If we consider the condition of vanishing torque on the dipole, then
using the constraint $<\Pi_z>=<p_z>=0$,  we obtain the energy
eigenvalues
\begin{equation}\label{e9}
E_\nu^{(\sigma)}=(\nu+\frac{1}{2}(1+\sigma))\mid\omega_{eff}\mid
\end{equation}
where $\nu=0,1,2,3.......$\\
The expression is similar to that of the standard Landau quantization, where charge $q$ in a uniform magnetic field
perpendicular to its motion has energy eigenvalues
\begin{equation}\label{h3}
E_\nu=(\nu+\frac{1}{2})\hbar |\omega|+\frac{\hbar^2 k_z^2}{2m}
\end{equation}
where $\nu=0,1,2,3.......$ and $k_z$ are real valued\\
In the standard Landau problem, the motion in the x-y plane is transformed into one -dimensional harmonic oscillator accompanying free motion in the z-direction and energies are independent of direction of revolution.
It is noted that in the atomic analogue case the energy eigen values depends on $\sigma$, the direction of revolution, and the condition of vanishing torque on the dipole puts an additonal constarint on the stationary states.

We can also make a comparison on the separation between two energy levels in both the cases.
   The
separation between two Landau levels is given by \cite{1}
\begin{equation}\label{e10}
\Delta E=\hbar|\omega|=\hbar\frac{|qB|}{m}
\end{equation}
with cyclotron frequency $|\omega|= \frac{|qB|}{m}$.\\
In our case, from eqn.(\ref{e9}) and (10)we find that the separation between two energy levels is
\begin{equation}\label{e11}
\Delta
E=\hbar|\omega_{eff}|=\hbar\frac{\mid{d_{m}\frac{\rho_{0}}{\epsilon_0}}+d_{e}{(\frac{\partial
B_{z}}{\partial x})\mid}}{m}
\end{equation}

With the help of the relation $B=\frac{\Phi}{S}$, ($S$ being the area and $\Phi$
the enclosed flux) and comparison of  eqn.(\ref{e11}) and
eqn.(\ref{e10}), we can find the analogy
\begin{equation}\label{e12}
q\Phi\longleftrightarrow
{\mid{d_{m}\frac{\lambda}{\epsilon_0}}+d_{e} S{(\frac{\partial
B_{z}}{\partial x})\mid}}
\end{equation}
where $\lambda={\rho_0}{S}$, is uniform linear charge density. This
relation shows the duality between the standard
Landau problem and a system of atoms with
permanent electric dipole moment and non vanishing magnetic moment
interacting with electric field and inhomogeneous magnetic field.

\section{Motion in symmetric gauge}

We may now proceed to solve the problem in a specific gauge, for example: symmetric gauge, and derive the energy spectrum.

The Hamiltonian (\ref{h1}) may be rewritten as
\begin{equation}\label{h2}
    H= \frac{{\bf (p-A_{eff})^2}}{2m}
    +\frac{d_m }{2m}{\bf \nabla}.{\bf E}+ \frac{d_e }{2m} |(\bf
n^\prime . \nabla){\bf B}|
\end{equation}
 Substituting the symmetric gauge ${\bf
 A}_{eff}=(-\frac{B_{eff}}{2}y,\frac{B_{eff}}{2}x)$ and considering the
 charge density producing the electric field to be
constant and specifying a constant gradient of the magnetic field
component (i.e. $\frac{\partial B_z}{\partial x}=$constant) the
Hamiltonian in planer motion become
 \begin{equation}\label{h4}
 \begin{array}{lcl}
 H&=&\frac{1}{2m}\left((p_x-A^x_{eff})^2+(p_y-A^y_{eff})^2\right)\\&&\\
&=&\frac{1}{2m}\left((p_x+\frac{B_{eff}}{2}y)^2+(p_y-\frac{B_{eff}}{2}x
)^2\right)\\&&\\ &=&\frac{1}{2}(p_x^2
+p_y^2)+\frac{1}{2}m\left(\frac{B_{eff}}{2m}\right)^2(x^2+y^2)-\frac{B_{eff}}{2m}L
\end{array}
\end{equation}
where $L=xp_y-yp_x$. To solve H, we change the variables to
\begin{equation}\label{h5}
z=x+i y,~~~~~~~~~~~~p_z=\frac{1}{2}(p_x-i p_y)
\end{equation}
and introduce two sets of annihilation and creation operators
\begin{equation}\label{h6}
    a=\frac{1}{2}\sqrt{\frac{B_{eff}}{2}}\bar{z}+i\sqrt{\frac{B_{eff}}{2}}p_z~~~~~~~
    a^\dag=\frac{1}{2}\sqrt{\frac{B_{eff}}{2}}{z}-i\sqrt{\frac{B_{eff}}{2}}\bar{p_z}
\end{equation}
\begin{equation}\label{h7}
    b=\frac{1}{2}\sqrt{\frac{B_{eff}}{2}}{z}+i\sqrt{\frac{B_{eff}}{2}}\bar{p_z}~~~~~~~
    b^\dag=\frac{1}{2}\sqrt{\frac{B_{eff}}{2}}\bar{z}-i\sqrt{\frac{B_{eff}}{2}}{p_z}
\end{equation}
satisfying the following commutation relations
\begin{equation}\label{h8}
    [a,a^\dag]=[b,b^\dag]=1
\end{equation}

We can write the $L$ and $H$ in terms of these operators
\begin{eqnarray}
  L &=&(a^\dag a-b^\dag b) \\
  H &=& \frac{B_{eff}}{2m}(a^\dag a+b^\dag b+1)- \frac{B_{eff}}{2m}(a^\dag a-b^\dag b) \\
   &=& \frac{B_{eff}}{m}(b^\dag b+\frac{1}{2})
\end{eqnarray}

Thus the motion is similar to that of a one dimensional harmonic oscillator.
So the eigenfuncion and energy spectrum of our system is given by
\begin{equation}
    \psi_\nu=\frac{1}{\sqrt{(2m\omega_{eff})^\nu \nu!}}(b^\dag)^\nu |0>,
    \end{equation}

\begin{equation}
    E=\frac{B_{eff}}{m}(\nu+1/2)=\frac{\omega_{eff}}{2}(2\nu+1),~~~~~~\nu=0,1,2,....
\end{equation}
respectively.

\section{Motion in noncommutating plane}

We may now study the specific system in noncommutating (NC) plane. The space
coordinate is noncommutating following the algebra
\begin{equation}\label{h9}
[\tilde{x}, \tilde{y}]=i\theta
\end{equation}
where $\theta $ is a constant noncommutative parameter.

Noncommutating coordinates can be written in terms of commutating
$(x,y)$ as \cite{non, nc1}
\begin{equation}\label{h10}
\tilde{x}\equiv
x-\frac{\theta}{2}p_y~~~~~~~~~~~~~~~~~\tilde{y}\equiv
y+\frac{\theta}{2}p_x
\end{equation}

So NC analogue of Hamiltonian (\ref{h4}) will be reduced to
\begin{equation}\label{h11}
\begin{array}{lcl}
    H_{nc}&=&\frac{1}{2m}\left([1+k]p_x+\frac{B_{eff}}{2}y\right)^2
    +\frac{1}{2m}\left([1+k]p_y-\frac{B_{eff}}{2}x
\right)^2\\&&\\&=&\frac{1}{2m}\left((\tilde{p}_x+\frac{B_{eff}}{2}y)^2+(\tilde{p}_y-\frac{B_{eff}}{2}x
)^2\right)
\end{array}
\end{equation}
where $$k=\frac{B_{eff}}{2}\theta~~,~~[1+k]p_i=\tilde{p}_i~~,~~\tilde{\omega}=[1+k]\omega~~\mathrm{and}~~\tilde{L}=x\tilde{p}_y-y\tilde{p}_x$$

To solve $H_{nc}$, we change the variables to
\begin{equation}\label{h5n}
z=x+i y,~~~~~~~~~~~~\tilde{p}_z=\frac{1}{2}(\tilde{p}_x-i \tilde{p}_y)
\end{equation}
and introduce two sets of annihilation and creation operators
\begin{equation}\label{h6n}
    \tilde{a}=\frac{1}{2}\sqrt{\frac{B_{eff}}{2}}\bar{z}+i\sqrt{\frac{B_{eff}}{2}}\tilde{p}_z~~~~~~~
    \tilde{a}^\dag=\frac{1}{2}\sqrt{\frac{B_{eff}}{2}}{z}-i\sqrt{\frac{B_{eff}}{2}}\bar{\tilde{p}_z}
\end{equation}
\begin{equation}\label{h7n}
  \tilde{b}=\frac{1}{2}\sqrt{\frac{B_{eff}}{2}}{z}+i\sqrt{\frac{B_{eff}}{2}}\bar{\tilde{p}_z}~~~~~~~
    \tilde{b}^\dag=\frac{1}{2}\sqrt{\frac{B_{eff}}{2}}\bar{z}-i\sqrt{\frac{B_{eff}}{2}}{\tilde{p}_z}
\end{equation}
satisfying the following commutation relations
\begin{equation}\label{h8n}
    [\tilde{a},\tilde{a}^\dag]=[\tilde{b},\tilde{b}^\dag]=1
\end{equation}

We can write the $\tilde{L}$ and $H_{nc}$ in terms of these operators
\begin{eqnarray}
  \tilde{L} &=&(\tilde{a}^\dag \tilde{a}-\tilde{b}^\dag \tilde{b}) \\
  H_{nc} &=& \frac{B_{eff}}{2m}(\tilde{a}^\dag \tilde{a}+\tilde{b}^\dag \tilde{b}+1)- \frac{B_{eff}}{2m}(\tilde{a}^\dag \tilde{a}-\tilde{b}^\dag \tilde{b}) \\
   &=& \frac{B_{eff}}{m}(\tilde{b}^\dag \tilde{b}+\frac{1}{2})
\end{eqnarray}
The Hamiltonian is now reduced to the form of a one dimensional
harmonic oscillator with frequency of precession
$\tilde{\omega}_{eff}$. So the eigenfuncion is given by
\begin{equation}
    \psi_\nu=\frac{1}{\sqrt{(2m\tilde{\omega}_{eff})^\nu \nu!}}(\tilde{b}^\dag)^\nu |0>
    \end{equation}
and the energy spectrum will be denoted as
\begin{equation}
    E_{nc}=\frac{\tilde{\omega}_{eff}}{2}(2\nu+1),~~~~~~\nu=0,1,2,...
\end{equation}

It is easily cheked that for $\theta=0$, i.e. for vanishing non-commutative parameter,
 the equations (40) and (41) are mapped to the equations of the eigenfunctions and eigenenergies of the commutative plane.

\section{Discussions}
We have studied  the quantum dynamics of a neutral
particle in two dimensions in external fields and found the analogy with the
standard Landau quantization. The neutral atom with permanent electric dipole moment and nonvanishing magnetic moment is considered in external electric
and  space dependent magnetic fields. The resulting
effect is analogous to that of a charged particle moving in presence of
a uniform transverse magnetic field. Our analysis is a more generalized version
of the Landau-Aharonov-Casher quantization \cite{eric} where the
atomic motion was considered in presence of electric field only. We have taken
nonvanishing gradient of magnetic field by considering a varying
magnitude of the z-component of the field along x-direction. The
same effect may also be obtained by varying the direction of the
dipole \cite{pach}.
The eigenfunction and energy spectrum of the system is derived using symmetric gauge
 and extended our analysis to study  the system in noncommutating plane. It is found that
the result reduces to that of the commutating plane when the noncommutative parameter
is set equal to zero.

Acknowledgement: Thanks to the referee for his fruitful suggestions.

\end{document}